# The Simulated Sky: Stellarium for Cultural Astronomy Research

Georg Zotti

Ludwig Boltzmann Institute for Archaeological Prospection and Virtual Archaeology, Vienna, Austria
Georg.Zotti@univie.ac.at

Susanne M. Hoffmann

Friedrich-Schiller-Universität Jena, Michael-Stifel-Center/ Institut für Informatik and Physikalisch-Astronomische Fakultät, Jena, Germany
susanne.hoffmann@uni-jena.de

Alexander Wolf

Altai State Pedagogical University, Barnaul, Russia
alex.v.wolf@gmail.com

Fabien Chéreau

Stellarium Labs, Toulouse, France
fabien.chereau@gmail.com

Guillaume Chéreau

Noctua Software, Hong Kong
Guillaume.chereau@gmail.com

**Abstract:** For centuries, the rich nocturnal environment of the starry sky could be modelled only by analogue tools such as paper planispheres, atlases, globes and numerical tables. The immersive sky simulator of the twentieth century, the optomechanical planetarium, provided new ways for representing and teaching about the sky, but the high construction and running costs meant that they have not become common. However, in recent decades, "desktop planetarium programs" running on personal computers have gained wide attention. Modern incarnations are immensely versatile tools, mostly targeted towards the community of amateur astronomers and for knowledge transfer in transdisciplinary research. Cultural astronomers also value the possibilities they give of simulating the skies of past times or other cultures. With this paper, we provide an extended presentation of the open-source project Stellarium, which in the last few years has been enriched with capabilities for cultural astronomy research not found in similar, commercial alternatives.

**Introduction**

Artificial illumination has utterly changed the human experience of the nocturnal sky in large parts of the world. As much as 80% of North Americans and 60% of Europeans cannot see the Milky Way from their homes (Falchi *et al.* 2016), and the current generation of children living in cities and towns may never have experienced a natural (or at least a largely natural) night sky dominated by the light of the Moon and stars and the sublime glow of the Milky Way, which since ancient times has evoked the proverbial "Ultimate Question of Life, the Universe and Everything" (Adams 1979). The extent to which an awareness of the skyscape and its past cultural significance has diminished can be seen in frequent questions at public stargazing events along the lines of "What? You can really see the planets without a telescope?" Yes, you can – and this is what has inspired many philosophers and early scientists (Starlight Initiative 2007). At the same time, however, those areas of the world most affected by light pollution are where computer screens, mobile phones and other electronic gadgets offer broad educational potentialities for educational simulations of nature, including the night sky. While no present simulation can completely recreate the natural outdoor experience with its solemn nocturnal impressiveness, at least some visual components can be replayed and can hopefully keep a principal interest in the sky awake.

Simulated skyscapes are also crucial tools for researchers in the fields of history of astronomy, archaeoastronomy and ethnoastronomy when attempting to interpret the sky in combination with human-made structures, for example floor plans of excavated buildings and monuments or remains of former cave dwellings. Over centuries and millennia, both the slight change in tilt, and to a greater extent, precessional movement of the Earth's axis have changed the orientation of the sky relative to the horizon. In addition, the proper motion of stars has significantly changed the position of a handful of bright ones since their place was catalogued in antiquity. This means that even under the best conditions without light pollution, today's natural sky over old structures does not provide the same view the builders of these structures would have seen with their own eyes.

The open-source desktop planetarium software Stellarium was created in response to problems such as these, and in recent years it has gained great significance in the field of cultural astronomy. Most readers may already be familiar with it and or use it in their research, including to illustrate papers and books – which we as its creators welcome, especially when properly acknowledged. In the *Journal of Skyscape Archaeology* we have so far published only a few brief "Software Notes" (Zotti and Wolf 2017; 2018), and in this paper we want to provide more detailed insights about how and why Stellarium was created, what it can be used for (as of version 0.20.4 – see Stellarium 2020), how accurate we consider it to be and how users can help to improve and further extend it. For context, we begin with a brief overview of historical astronomical technology, after which we present a brief history of the project and its uses for virtual archaeoastronomy, which also drove much of the project's recent efforts for increased accuracy. We then discuss the application of exchangeable "skycultures", a feature which was originally geared mostly towards use in local communities but is attracting increasing interest by researchers in ethnoastronomy and even astrophysics. We then conclude with a brief discussion.

**Learning the Constellations: A Short History of Simulated Skyscapes**

From the seventeenth to the twentieth century, astronomers working in the Western tradition were familiar with the application of planispheric star maps made from cardboard, the simplified descendants of the medieval planispheric astrolabe. These interactive maps make it possible to learn the constellations, identify what stars would be visible on a particular date, or even find a good estimate for rising or setting times, by aligning a star map surrounded by a circularly enclosing date scale with a horizon mask that shows a circularly enclosing time scale. They are easy to transport but have disadvantages: for example, they are only computed for a particular geographical latitude, the part of the sky revealed by the horizon mask appears heavily distorted and the planet positions are not displayed but have to be taken from a yearly almanac. A more luxurious instrument, which was mostly confined to indoor study, was the celestial globe (Mokre 2008; Hoffmann 2017), the concept of which had been described by Ptolemy (*Almagest* 8.3–6) in antiquity. Some globes were large, such as the 4 m Coronelli globes created for Louis XIV (Coronelli 1683; Milanesi 2018) and the Globe of Gottorf, built in 1664 (Meier 1992; Lühning 1997), and there were also great walk-in globes, such as the now-lost "Pancosmos" of Erhard Weigel that was installed at the city palace of Jena (Jenaer Stadtschloss) in 1661 (Meier 1992; Kratochwil 2011), and later examples that could be experienced by the public at the World's Fairs in London in 1851 and Paris in 1889 (Geppert 2010).

    The twentieth century saw the development of an immersive environment in which the sky and the movements of its various features – Sun, Moon, planets, fixed stars and so on – could be experienced by urban populations. This was the optomechanical projection planetarium (for example, Firniss 1980; Meier 1992; Boyce-Jacino 2018), first developed in the 1920s. This system simulates a view of the upper hemisphere of our environment projected onto a dome. A variety of projection systems was developed during subsequent decades, progressing from tilted domes to omnitheatres with auxiliary projectors at the periphery and domes that dispense with a central projector and use video projectors at various angles. In their attempts to produce an almost perfect simulation of the real sky, all systems have their pros and cons (Rienow 2013 and references therein), but sky simulation above the mathematical horizon is reasonably complete in modern projection planetariums. In early installations, the local horizon was usually modelled with a cardboard or sheet metal skyline silhouette of the respective city, with dim lights shining from behind around the horizon even simulating the horizon brightening caused by urban light pollution. In the 1980s, the development of new projection systems allowed these fixed panoramas to be replaced by images cast by auxiliary projectors.

An advantage of the dome planetarium is the way that the immersive experience makes science accessible to wide audiences. This includes historical astronomy: a popular show of the 1970s in Vienna's planetarium (and many other planetaria) was a presentation about the "Star of Bethlehem" and its interpretation by Ferrari d'Occhieppo (1969; inspired by Johannes Kepler) as a triple conjunction of Jupiter and Saturn as they met three times within a few months in 7 BC. The mechanical planetarium projector had been found to be accurate enough for this kind of presentation (Mucke 1967), but it required many hours of supervised backward winding of the planetary gears. A more significant limitation is that planetarium domes are scarce and rarely available for pure research, both due to the costs involved – planetariums are usually private enterprises – and the training required to work with them. Also, there is a drawback with the classical hemispherical dome and a projector optimised to keep spherical angles correct when projected onto it: a foreground which could provide archaeological contexts such as the foundations of a prehistoric monument cannot be displayed on a domed screen that cuts off the view at the mathematical horizon (Zotti *et al.* 2006). Only the latest generation of digital planetaria can create the required foreground context by digitally contracting the sky and projecting a field of view larger than 180° onto the hemispherical screen, so that enough of the ground becomes visible along the edge of the dome and still aligns correctly with the – now intentionally distorted – celestial geometry.

The early era of home computers in the 1970s and 1980s allowed the development of software for astronomy enthusiasts (Meeus 1988; 1998; Montenbruck and Pfleger 1989). First limited to numerical data output, the arrival of interactive graphics output, much of which was driven by the computer games industry, also allowed the development of various star-mapping programs. Popular astronomy magazines provided programming columns, and finally the World Wide Web simplified the distribution of programs and data. By the late 1990s, these programs provided a fairly realistic representation of the sky which could be set for any location on Earth or even on other planets. Some allowed the representation of mountains or houses along the local horizon to intensify the sense of immersion, first by use of polygonal obstruction masks and later by a panorama photograph. Many such programs represented the sky as a hemispherical view in stereographic projection known from earlier printed material, while others used the perspective view more common to photographic and video cameras and computer games. This view makes it possible to zoom in on particular objects, whereas very wide-angle perspective views are extremely distorted and can never show the full sky. Such programs, also based on improved models of planetary motions like VSOP87 (Bretagnon and Francou 1988), finally provided views and a simulation accuracy approaching, or even in some ways surpassing, the quality of the projection planetarium (although lacking the immersive quality of the dome) and thus invited their application in historical research which had hitherto been based on numerical tables (for example, Hunger 1994; 1997).

In recent years, the mostly portable telescopes used by amateur astronomers have been augmented with computer control for locating and tracking objects, and some of today's amateurs can now create photographs that regularly surpass the best professional results achieved with much larger instruments only a few decades ago. Computer soft-ware, often included as components of desktop planetarium programs, can aid in setting and controlling telescopes and instrumentation. For observation planning, and during observation in the field, computer desktop planetarium programs, and more recently also programs running on smartphones and tablet computers, have largely superseded the classical cardboard planisphere and often also the classical printed sky atlas.

**A Brief History of Stellarium**

In the summer of 2000, one of us (FC) as a student started a project that was both a hobby and connected with his studies. The aim of his "Stellarium" was to provide a realistic simulation of the sky, using state-of-the-art 3D computer graphics technology. At that time, OpenGL, a programming library for real-time computer graphics, had started to be supported by affordable consumer graphics cards on the three major desktop platforms of Linux/X11, Microsoft Windows and Apple Macintosh. He soon decided to make the program's source code publicly available under the GNU Public License and invited fellow developers to join the project (Stellarium 2020). The project then started to take off, especially among astronomy enthusiasts and planetarium owners, and by 2006 a team of four to five volunteer developers distributed around the globe was very active on the project. Each member contributed specialist expertise, bringing in major features such as large star catalogues, various sky projections, improved planetary accuracy, a scripting feature and a modular plugin (extension) system, as well as a revamped GUI based on the Qt programming framework (Qt 2020) and a new website and user guide. The project then quickly gained visibility, grew its user base to an estimated 10–15 million users and won several awards such as being elected "Project of the Month" on sourceforge.net in May 2006 (Stellarium 2006). In 2007, Fabien started to work full-time on Stellarium as part of his work at the European Southern Observatory (ESO), where he extended Stellarium to enable browsing and displaying professional astronomical observations made by large observatories such as the Very Large Telescope (VLT – Kapadia *et al.* 2008). This sponsorship also permitted the costly refactoring without which such quickly growing C++ projects often become impossible to maintain.

By 2010, this first team of developers had developed Stellarium into a stable program with an unconventional but intuitive user interface that surpassed the visual quality of most other astronomical simulators. It could control several types of computerised telescopes, simulate the view through ocular equipment and identify the locations of artificial satellites. Most user complaints had to do with graphical issues, and in most of these cases users had not installed the appropriate graphics drivers. However, numerical accuracy was also criticised, especially when simulating historical skies, and indeed, a chain of improvements on this issue has been made in the versions that have since appeared.

One major leap forward was support for ΔT, the difference between Terrestrial Time (TT) and Universal Time (UT) caused by the irregular slowdown of Earth's rotation, introduced in version 0.12.0 (2013) – to our knowledge, Stellarium is the only program so far to even include support for many models of ΔT, which allows users or researchers to visualise the effects of the application of particular models or to research a custom model by assigning the coefficients for the common quadratic formula to compute ΔT. Adding the computation of ΔT greatly improved the accuracy of visualisation for solar eclipses in the program. However, ΔT is known only from solar eclipse records that date back to the first millennium BC, so that a simulation of solar eclipses, and any conclusions that are based on a "virtual observation" of a solar eclipse from a particular location at some much earlier date, should be regarded with much caution. This uncertainty of ΔT is not a problem for Stellarium in particular, but for all known simulations.

The other major improvements in accuracy were the adoption of a long-time model for the precessional motion of Earth's axis (Vondrák *et al.* 2011; 2012) in version 0.14.2 (2015) and access to the DE430/431 ephemerides (Folkner *et al.* 2014) since version 0.15 (2016).

A major internal change was the upgrade of Qt (a C++ programming library and the framework for cross-platform development of the user interfaces used by Stellarium), which reflects Stellarium's step from version 0.12 (the last series based on Qt4) to 0.13 (based on Qt5). Many users were still running old hardware and demanded many more years of support for the technically obsolete 0.12 series. Unfortunately, after the switch to Qt5 most of the original team members decided to leave the project, and since around 2014 Stellarium has mostly been developed further by a team of only two (AW, GZ), with major contributions in functionality developed in collaboration with students from TU Wien (Vienna University of Technology – Zotti 2016; Zotti *et al.* 2017), other students sponsored by the European Space Agency's Summer of Code in Space programme and, mostly for various telescope control features, by several other external contributors. Additionally, operating and development systems have continued to evolve and years after Windows XP had been officially retired, Stellarium also had to stop supporting its use on that system. We will likely face a similar upgrade challenge to Qt6 next year (2021), but cannot currently say in detail what will change.

An Astronomical Calculations (AstroCalc) window was introduced in version 0.15.0 (2016) as a simple tool to compute planetary phenomena and ephemerides in tabular form and also as plotted curves in the sky. This tool has been greatly enhanced in successive versions. It is aimed at amateurs, for planning their own observations, and at educators and researchers who wish to obtain some quick advanced computations, such as relations between magnitude (brightness) and phase angle for planets over several years (simultaneous visualisation of two functions of time in a separate panel), the horizontal positions of the Sun observed on any Solar system object during a year (visualisation of the analemma) or a plot of the path of a comet on the sky (ephemeris).

### *Spinoffs, Forks and Related Projects*

Nowadays, the name "Stellarium" is used in several contexts. The desktop version (on which this paper is focused) is currently the most complete system. However, by 2006 its high visual appeal was already sufficient for use of the program in small planetarium domes. One early contributor separated and "forked" a project called NightShade, from which later Stellarium360 (2012) and SpaceCrafter (2016) were derived. Another digital planetarium fork is ShiraPlayer (2013). All of these have been developed with the particular needs for planetarium shows in mind (for example, a dome video player). However, each fork means leaving the main developing line, and new features found in Stellarium cannot be expected to be available automatically in the spinoff projects and *vice versa*. The upcoming era of smartphones invited another official fork by one of us, "Stellarium Mobile", which is based on version 0.13 and runs on most Android smartphones.

Further development of network infrastructure and the ubiquity of internet access invited the creation, in 2017, of an entirely new project, Stellarium Web, initiated by Guillaume Chéreau. The goal of this project was to create a new version of Stellarium, light enough to run directly in a web browser. This was achieved by re-coding the low-level part of the engine from scratch, almost exclusively in pure C language without any external dependencies, and converting it into a JavaScript library using the *emscripten* compiler (emscripten 2020). The new engine was also designed to handle large amounts of online imaging and sky object catalogues instead of relying on local file-based data. The version currently online (Stellarium Web 2020) has a binary size of less than 2 MB (allowing fast download and start-up), runs at 60 fps (frames per second) on most recent computers and can display most celestial objects with a magnitude brighter than 20. However, despite these exciting capabilities, this web version still has only a limited set of features and the accuracy of planetary computations for times in the distant past or future is not yet on a par with its desktop counterpart. It is therefore not recommended for use in archaeoastronomy, although this may change in the future.

**Stellarium for Cultural Astronomy**

On the surface, archaeoastronomy combines the skills of both archaeologists and astronomers – researchers dealing with material evidence of past human cultures on the ground or buried in the soil versus those observing and studying celestial phenomena and, more recently, the underlying physics. One important field in archaeoastronomy is concerned with the orientation of architecture, like temple axes or building entrances, towards particular phenomena observed in past skies, such as solstitial sunrises or sunsets, extreme points in the path of the Moon or the rising and setting points of particularly bright stars. Most aspects of this research, and also most studies about the astronomies of other cultures, require only classical positional astronomy and rarely deal with the physical nature of celestial objects.

An exception is the question of the nature of "transients" (short-lived unexpected celestial phenomena, e.g. comets, or stars flaring up as novae or supernovae) among historical observations – a field which builds a bridge between historical astronomy and astrophysics. As there are individual data points some hundred or even some thousand years ago, these observations could possibly contribute to the research on and modelling of the long-term evolution of close binaries such as symbiotic stars, cataclysmic variables, supernova remnants, black holes, star mergers and other objects from high-energy astrophysics. Therefore, the correct and proper simulation of past views of the sky and of the various sky cultures is crucial for these fields of research (see below).

*Virtual Archaeology*

Many archaeologists nowadays use CAD (Computer-Aided Design) or GIS (Geographical Information System) software to document their sites. Survey points measured in the field are stored in a database and can be displayed together with outlines of features identified during excavation or even in images created from geophysical prospection in digital maps. The terrain, all-important for horizon and orientation studies, is now surveyed with airborne laser scanning (LiDAR), from which high-resolution digital surface models (DSM) and digital elevation models (DEM) can be computed. For the latter, vegetation and built structures can be filtered out from the data quite easily, which also allows a detailed view even under tree canopies. These surveys at least provide data about present-day topography, and larger or human-made changes or the effects of natural slow erosion or landslides can partially be modelled backwards. Further, simple 3D models of past architecture can be created as visualisation aids just from vertically extruded feature polygons. Teams which include 3D artists and architects can recreate whole cities from antiquity on the screen. "Game engines" – software frameworks for 3D computer game technology – allow the creation of lively scenes enriched with sounds, animated objects or creatures in virtual environments that can be explored interactively. To keep the displayed reconstruction in line with scientific knowledge and to separate scientific knowledge clearly from hypothetical reconstruction (which may also include reconstruction variants for discussion), standards like the London Charter (Denard 2009) and the Seville Principles for Virtual Archaeology (Carrillo Gea *et al*. 2013) have been established.

Most architectural computer reconstructions of the past are shown under a daylight sky. Rendering can be controlled with sunlight that can be configured to any location in the sky. When the model's geographic location is known to the rendering system, a sky module may be able to compute the Sun's position from a calendar interface. This is useful for contemporary architectural simulation, to analyse the impact of a planned building's shadow on its vicinity, but is frequently limited to dates close to the present time.

Unfortunately, most architectural simulation systems or game engines have no built-in support for the current night sky and even less for that of past times: at best they simply show a static "skybox" with the starry night sky at some unspecified time. The slight variation of solstitial points or a rotating starry sky which undergoes precessional motion has to be created by the modelling team.

In the last few years Stellarium has therefore been extended with unique features geared towards research and demonstrations in *Virtual Archaeoastronomy*.

*ArchaeoLines*

Stellarium's functionalities can be extended with optional modules, called plugins. One of these plugins, ArchaeoLines, shows the diurnal tracks or, more accurately, the declinations of the Sun on the dates most often considered as important for archaeoastronomy:

the Sun's paths at the solstices, equinoxes and the "cross-quarter" dates. These "cross-quarter" dates still exist as alternative season markers in our calendar. If we re-define the seasons to not begin with equinoxes and solstices, but instead to be centred on those key dates, in terms of cultural festivals in the West spring begins on Candlemas (2nd February), summer begins on May Day (1st May) and winter on All Saints' Day/Samhain (1st November). A line in the sky indicating the path of the Sun on those dates is defined in the program by the Sun reaching the following ecliptical longitudes:

$$\lambda = 45° + n \times 90°, n \in [0,1,2,3]$$

This can simplify the process of finding a correlation between architecture axes and sunrise or sunset points in the horizon. In a similar way, the Moon exhibits extreme points in its orbit at the declinations of major and minor standstills. These are the outermost and innermost turning points of the monthly declination range of the Moon. The Moon reaches its extreme declinations and outermost rising or setting points that represent the intersection of declination arc with the visible horizon with a periodicity of about 18.6 years, whereas it passes the declination of the innermost turning points basically every month. The displayed arcs highlight the maximum declinations the lunar centre can have at the mathematical horizon. The arcs are actually shown twice: the Moon is so close that the difference in distance when at apogee (farthest point from Earth's centre) or perigee (closest point) causes a noticeable difference in its horizontal parallax. It should also be noted that the declination of the Moon as it transits the meridian during the major standstill day may exceed the displayed arc slightly, caused by the effect of diurnal parallax.

The plugin can also show the diurnal track of the current declinations for a planet, any selected object or arbitrary declinations and azimuths. Also, declination arcs of zenith and nadir transit and the current azimuth and hour angle for the selected object can be displayed. To mark the direction towards a geographical target like a sacred place, the azimuth, or vertical arc into the direction of two such places, can also be displayed.

### The Landscape Horizon

Computing rising and setting points of celestial objects along the mathematical horizon is not difficult, but is rather pointless when we are investigating a site which does not have a flat horizon. In the northern hemisphere, every hill or mountain visible along the horizon shifts the apparent rising or setting azimuth of any celestial object towards the south. An archaeoastronomical site survey must therefore *crucially* take care to record the horizon altitudes into the relevant view directions (Figure 1). Stellarium can show polygonal lines along the horizon derived from such measurements, or display horizon images which optimally have been recorded on the spot of interest and carefully assembled into panorama photographs. It is also possible to overlay the polygonal line derived from measurement over the photo-based panorama. How to create such panoramas by stitching photographs taken from a single point of view with the free and open-source Hugin panorama maker (Hugin 2019) is documented in the Stellarium User Guide (Zotti and Wolf 2020b).

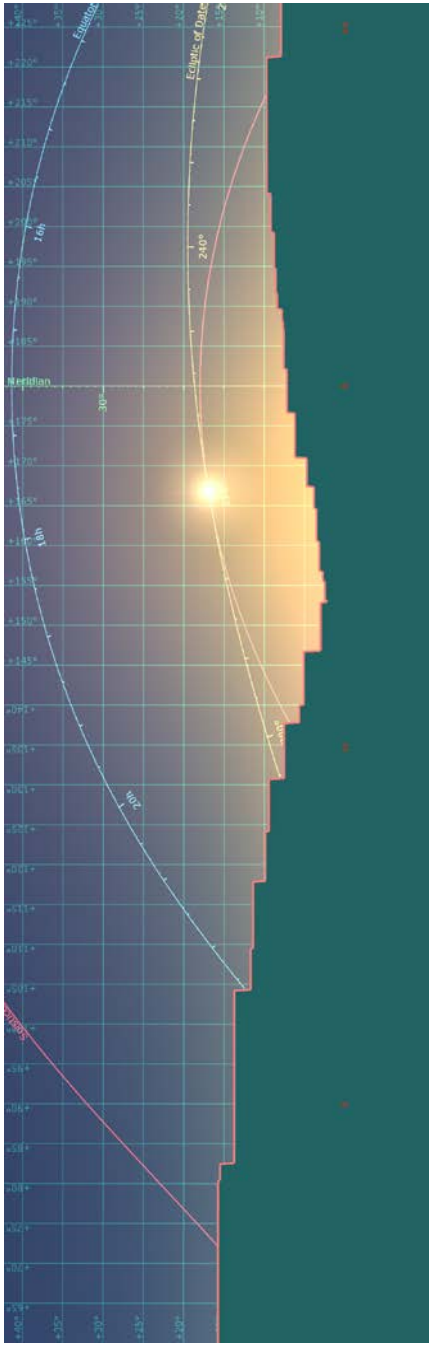 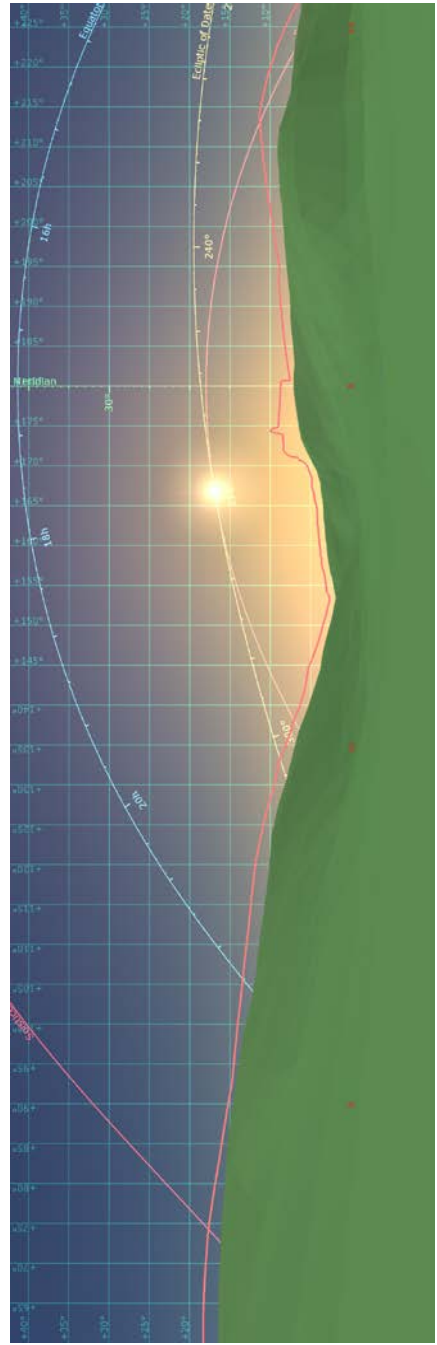

a  b

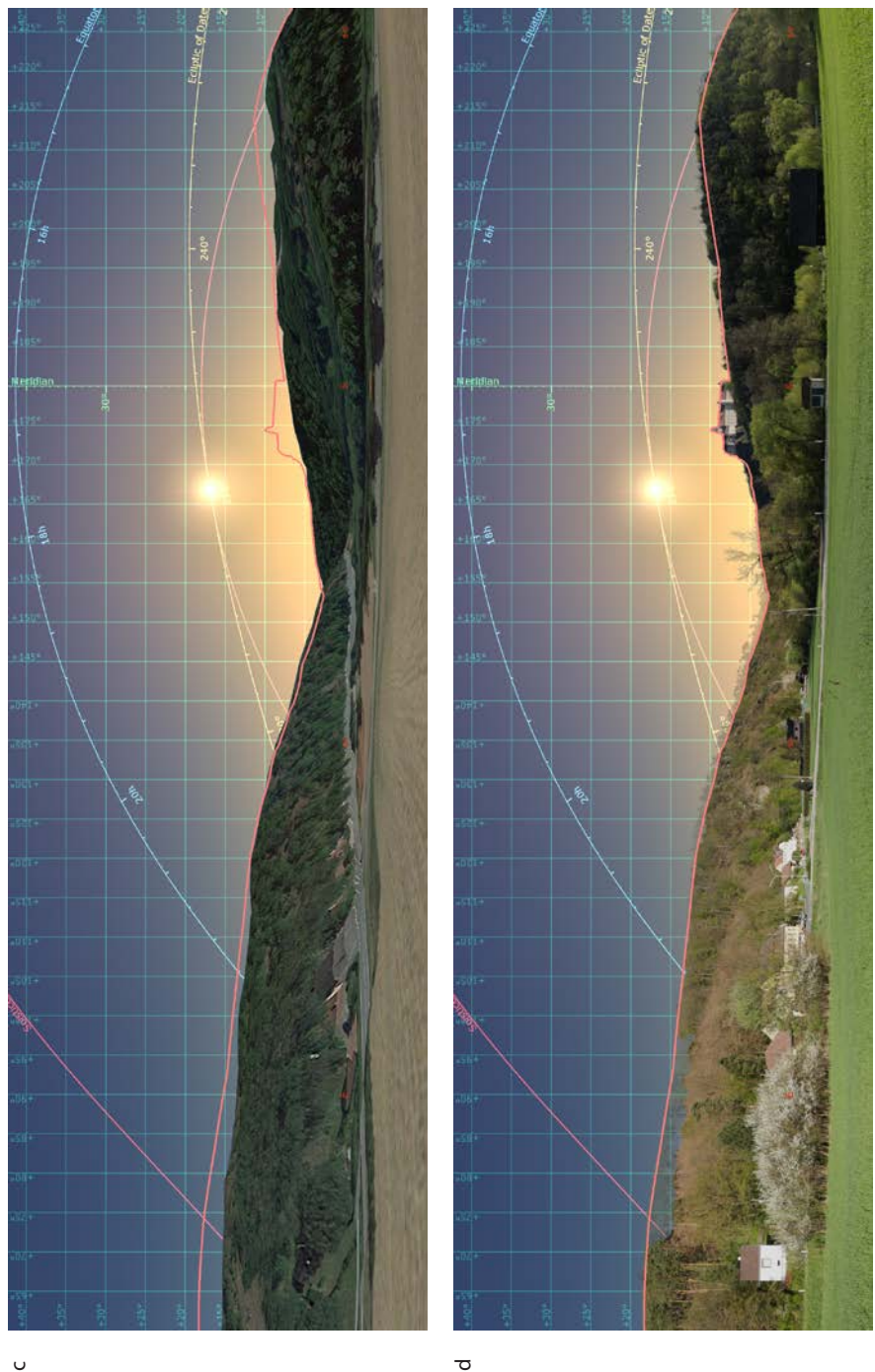

**FIGURE 1.** Comparison of artificial vs. measured horizons or "landscapes" in Stellarium for the site of the Neolithic circular enclosure near Rosenburg, Lower Austria. From top: (a) created by HeyWhatsThat – HoriZONE provides the same shape, but with more complete location data; (b) created with Horizon 0.13c from SRTM1" data; (c) created from artificial photo-like screenshots in Google Earth – note the contrast with SRTM1", especially with the hill shape near the right image edge (image copyright: Google); (d) photo-based panorama created during the ASTROSIM project (Zotti and Neubauer 2015) – the polygonal line represents measurements with a total station, which helped adjust the panorama and was copied into solutions b and c for better comparison. The measured line, however, is influenced by vegetation and should be expected to show differences from the artificial solutions.

In recent years a new kind of investigation, especially for preliminary work, has been established which can best be performed before actual site visits, and is based on digital data only. This is concerned with the application of GIS software to create the horizon polygon that defines the visual border between sky and ground. In addition, there are dedicated web services which can deliver such data based on SRTM3" (~90m) or even SRTM1" (~30m) digital surface models. One example is HeyWhatsThat (Kosowsky 2014), which also provides a gazetteer to identify the names of mountain peaks surrounding a site. Another dedicated web service named HoriZONE appeared recently which converts the HeyWhatsThat results into a polygonal landscape for Stellarium, allowing the addition of more information such as author, description or timezone in the process (Doyle 2019). Since summer 2020 HeyWhatsThat has provided its own direct download option for Stellarium landscapes (without the addition of these details). These horizon polygons define which parts of the sky are covered by some terrestrial feature, but do not reveal the distance to – or details in the appearance of – the horizon.

A visually more appealing option to create an artificial landscape horizon has been presented by Andrew Smith. His program Horizon (Smith 2020) can compute a landscape panorama from SRTM (3" or 1") and other DSM or DEM raster formats, and optionally also augment it with the diurnal tracks also found in the ArchaeoLines plugin mentioned above. Its internal viewer reveals the distance to a location under the mouse pointer. The artificial rendering can also be configured to give visual hints on elevation (by colour-coding) and distance (by reducing colour intensity) and the program can directly export a fully configured landscape package (terrain panorama rendering and a polygonal horizon line) for Stellarium.

Yet another option is provided by a workflow (Zotti 2013) that creates artificial panoramas from photo-like ground views in Google Earth (Google 2020). In some usually urban areas, the terrain is available with 3D details and even shows buildings and trees. It can be a valuable exercise to combine this manually with a polygonal landscape created by on-site measurement or one of the abovementioned methods to gain additional hints and insights about the appearance and quality of each method. In the optimal case, the polygonal line derived with one method should run along the seam between earth and sky visible in the picture derived with another method, unless the foreground is represented in higher detail.

SRTM is known to contain some elevation errors (see, for example, Ferranti 2014 and links from there), and the online services may use various solutions or even their own improved data sets to deal with such data errors or insufficient spatial resolutions. Each of the digital methods should be expected to show differences, especially in close-by areas. While these methods in general provide good background information for a distant horizon (when this is not perceivably damaged by a larger data error), in the near surroundings of our site of interest these SRTM-based DSMs are in general not accurate enough to represent possibly crucial detail. (Note that a 1 m vertical error causes a 1 arcminute lift or drop in over 3 km distance, and of course larger errors if closer.) The horizon line from HeyWhatsThat (Figure 1a) suffers most from lack of resolution in the SRTM data, which causes artificial steps that the other methods appear to interpolate.

There are also notable differences with respect to today's treeline. In Figure 1d, trees were painted semi-transparent in an area where glimpses of sky were visible through the trees, indicating lower ground which seems to be represented well in the Google Earth-based panorama (Figure 1c). The rendering of the Horizon program seems to be somewhat lower, at least in the vicinity of this site, but we cannot say whether this is caused by a difference in elevation data or by a processing error. Of note is the difference in the topography on the right foreground hill between Horizon/SRTM1" and Google Earth, indicating that elevation raster data used by Google Earth is different from the "official" SRTM. Therefore, currently we cannot give a clear preference or recommendation for one of these methods, apart from clearly stating which method and data were used. A panorama photograph taken on the spot of interest and carefully aligned with a surveyed horizon polygon, or at least adjusted to match the distant parts of an artificial panorama, usually provides higher detail and should reveal which parts of the scene appear to be important, at least when the archaeological monument in question is still at a minimum partially visible and when horizon altitudes formed by vegetation (treelines) have not changed significantly.

The inclusion of horizon panoramas in celestial simulation software is a valid approach to analysing the sky as observed from a few predefined viewpoints like temple corners, centres, main axes, entrances or similar well-defined spots. We may identify or confirm or disprove presupposed assumptions, but it is not easy to investigate further and address questions such as "what happens to visibility a few steps away from that spot?" Likewise, when we observe a potentially interesting solar alignment or shadow phenomenon during a short site visit today, we may ask ourselves "on which other days in the year does this phenomenon occur?" To answer these questions, we need at least three-dimensional mobility in our virtual location.

*Four-Dimensional Virtual Archaeoastronomy*

The most ambitious improvement towards developing modern tools useful for virtual archaeoastronomy was the addition of a 3D rendering plugin, Scenery3D (Zotti 2016), which allows interactive walkthroughs of virtual landscapes. Meanwhile, the model rendering module was even made time-aware (Zotti *et al.* 2018), so that georeferenced and time-annotated (phased) site reconstructions can be combined with the simulated skies of past millennia. The simulation always shows the astronomically correct sky over the reconstruction of a site at the selected date, and can also simulate shadows cast by the Sun, the Moon or even the planet Venus. Rendering of model components that do not fit the current date can be suppressed. The user can walk around in the virtual scene in first-person perspective with a configurable eye height above ground: if the scene is configured properly, the survey coordinates of the virtual eye point can be seen in the display and bookmarks can be set for interesting views. In the first-person perspective, building axes and the effects of light and shadow governed by windows or natural caves can be simulated in combination with what the sky would have looked like to the users of the original sites in question, making available views that cannot be experienced today even under the most pristine skies because the stars have shifted due to precession. For technical reasons, non-perspective rendering modes (for example, stereographic, fish-eye or cylindrical views) have to be rendered on six sides of a cube and then re-projected into the scene, which imposes a performance bottleneck for large models.

The rendering is suitable not only for scientific analysis, where a simple model may be sufficient, but is of a high quality that can also be used for architectural models geared towards public outreach. Small details in larger sculpted surfaces like walls decorated with inscriptions can be shown with *normal maps,* which can save computing operations. An average contemporaneous notebook PC can work with landscape models or models derived from laser scanning or image-based modelling with several millions of triangles. In the following examples we include performance values from one such notebook, an Acer Aspire V15 Nitro with Intel Core i7-6700HQ CPU, 16 GB RAM and Nvidia Geforce 960M with 4 GB dedicated GPU RAM, produced in 2016.

### *Example 1: Vienna Sterngarten*

The regular installation of Stellarium includes two 3D sceneries. One, aptly called "Test-Scene", is a little example of configuration options. The other is the model used while several phenomenological aspects of Stellarium were being developed and cross-checked: the "Sterngarten" (Star Garden) is a modern skywatching platform built 1997–2001 in the southwestern outskirts of Vienna by the late astronomy populariser Hermann Mucke (1935–2019) and the Austrian Astronomical Society (Mucke 2002). This public observing and teaching facility includes comprehensive architectural details for astronomical naked-eye phenomenology, especially as regards horizontal directions (azimuths) – a subject frequently discussed in archaeoastronomical contexts (Figure 2). It consists of a stepped pyramid platform which raises the observer over the local vegetation. The ideal observer should then be located standing in its centre. Most adult visitors should actually crouch slightly to match all indicated angles perfectly: a railing 1.5 m above the platform level forms the mathematical horizon. Two pillars to the north and south mark the meridian and bear altitude marks. Three plates on the southern pillar additionally indicate solar noon elevations for the solstices and equinoxes. Six outlying pillars indicate the geometrical azimuths for sunrise and sunset at the equinoxes and solstices on the mathematical horizon. (The western horizon is elevated, so that the real sunset will be seen earlier and farther south.) Short arms extending from each pillar along the mathematical horizon towards the north bear a notch which indicates the apparent location where first or last light of the Sun can be expected when refraction is taken into account. The outliers extend to 6° altitude, which can be mentally inverted to 6° depression where the Sun is positioned at the end (evening) or beginning (morning) of civil twilight, respectively.

The north pillar bears a disk with a hole, indicating the northern celestial pole. The disk itself covers the pole star to indicate that it is not exactly at the pole. To the north of the platform, the noon shadow of the pierced disk casts a diffuse spot of light inside an unsharp shadow ring onto a date scale along the local meridian. The shadow figure during the winter half of the year is cast onto an inclined mast, so that it is raised above any possible snow cover. This mast is parallel to the Earth's axis and simultaneously forms a giant gnomon for a sundial. One critical date is the equinox, when the disk's shadow falls on the foot point of the inclined mast. Figure 3 shows a simulation for this date, compared to a photograph. The shadow can be rendered unsharp by several methods found in the computer graphics literature. The most ambitious of those, PCSS (Fernando 2005), blurs shadows as a function of the distance from the shadow caster, allowing a fairly good simulation of the distance-dependent vanishing saliency of shadows cast by thin objects.

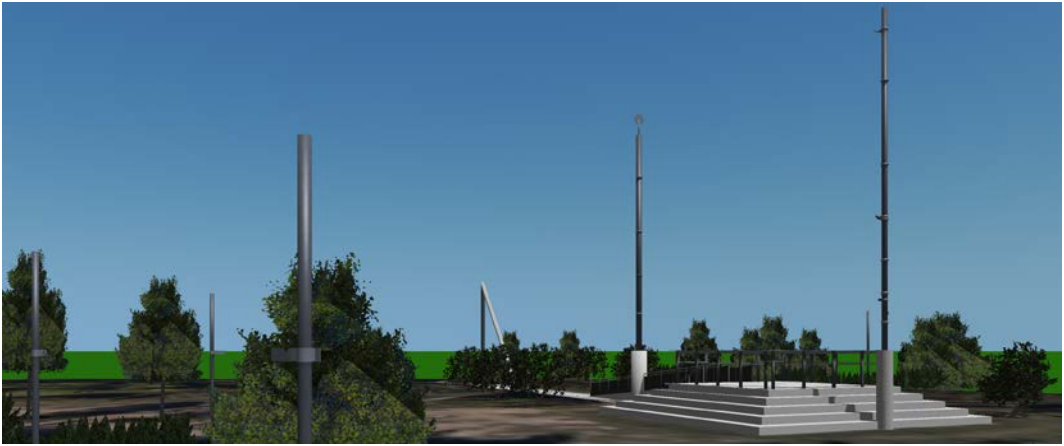

**FIGURE 2.** The "Sterngarten" modelled in Stellarium, viewed from the southwest. The real installation is surrounded by shrubs of a "suburban wilderness". In this display, the small terrain clip is surrounded by the "zero horizon" landscape, which only covers the ground below the mathematical horizon in green. The eye height in this view is configured to be 1.5 m above the platform centre, so that the railing and the side arms which extend from the outlying pillars are aligned with it.

The model was created on a small flat piece of terrain and a static landscape was exported before the 3D module was available to allow development and testing, and later demonstration of the effects of atmospheric refraction. Most other models should include more of the surrounding terrain and also a landscape panorama which should model the parts of the horizon which do not change perceivably when the observer moves within the central region of interest.

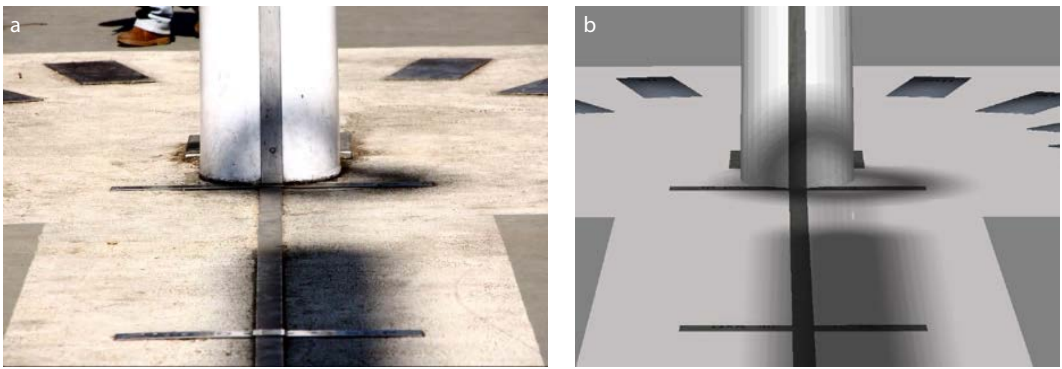

**FIGURE 3.** The meridian transit of the shadow figure of the disk on top of the central pillar marks local noon and the date. On equinox days, it transits at the foot of the inclined mast: (a) photograph taken by G. Zotti on 20th March, 2010; (b) faithful Stellarium representation of the solar shadow's blurred appearance, created by using the PCSS algorithm.

The Scenery3D plugin is slowly gaining attention, but while astronomical research data can usually be found online shortly after collection, high-resolution geodata, and more importantly archaeological and cultural heritage data, usually cannot just be downloaded, reconfigured and re-published in a different research context for free, but are stored away in the archives of its creators or their customers, or shown online in a format or with restrictive licensing that prevents download or further use. After the ASTROSIM project (Zotti and Neubauer 2012) and the study of the "Antinoeion" in Hadrian's Villa at Tivoli (Frischer *et al.* 2016), the first author (GZ) was asked for assistance in further projects which were welcome as stress tests with real-world data.

### *Example 2: Mnajdra Temple*

The prehistoric temples of Malta have been described extensively in the archaeoastronomical literature (see for example, Hoskin 2001; Ventura and Hoskin 2015). For an ongoing study (Lomsdalen 2014), laser scans of the Mnajdra temples, created before it was tented over to protect it from the elements, were provided by Heritage Malta, and aerial photography and laser scanning (LiDAR) data (digital elevation model, or DEM) by the Malta Planning Authority. Terrain analysis in ArcGIS 10.4 (Esri 2020) helped identify which part of the island was visible from the surroundings of the temple and therefore had to be included in the 3D model (Zotti 2019). From this part of the raster DEM a texturised triangulated irregular network (TIN) or "3D mesh" was created in ArcGIS 10.4 and exported with bespoke conversion tools into the widespread "Wavefront OBJ" format used by Stellarium's 3D rendering module. For technical reasons the vertex coordinates of the model cannot represent full UTM coordinate values (Universal Transverse Mercator, the currently most widespread survey coordinate system – Snyder 1987, 57–64), but must be shifted into a numerical domain appropriate for single-precision floating point arithmetic. The offset to the original UTM coordinates is stored in the scene configuration file.

The temple scans were converted from their obsolescent native data format and combined with the DEM mesh in the free and open-source 3D modelling program Blender (Blender 2020) into a single OBJ file which, together with the terrain model, consists of more than 10 million triangles which can be displayed at more than six frames per second in perspective rendering mode. However, the memory requirements for more than 1000 photo-based textures were too much for the laptop used for the test (the Acer Aspire V15 Nitro described above) and prevented shadow simulation, so a model with a monochromatic stone colour was created. The loss in visual quality – in this case of a limestone monument – should be bearable.

The user can move freely in this model and explore all rooms and cavities that were recorded by the laser scanning. We made no further attempt to clean up the scans or simplify the geometry, as further evaluation of this model is not our task. The question for us at this time was mostly as follows: how well does virtual archaeoastronomy with Stellarium work, and how large (in terms of model detail) can a model be handled and still remain useful? We cannot observe the sky of Malta's prehistory, and therefore we had to reformulate the question: how accurate is astronomical simulation of a highly detailed 3D model with Stellarium? This requires, of course, comparison with data taken during a site visit. This model was tested against photographs taken during an early-morning visit which formed an excursion highlight of the 2014 European Society for Astronomy in Culture conference (SEAC 2014). The first rays of the rising equinox Sun, when it rises over a little ridge east of the temple, enter the central axis of the southern temple (Figure 4, with central sections enlarged in Figure 5).

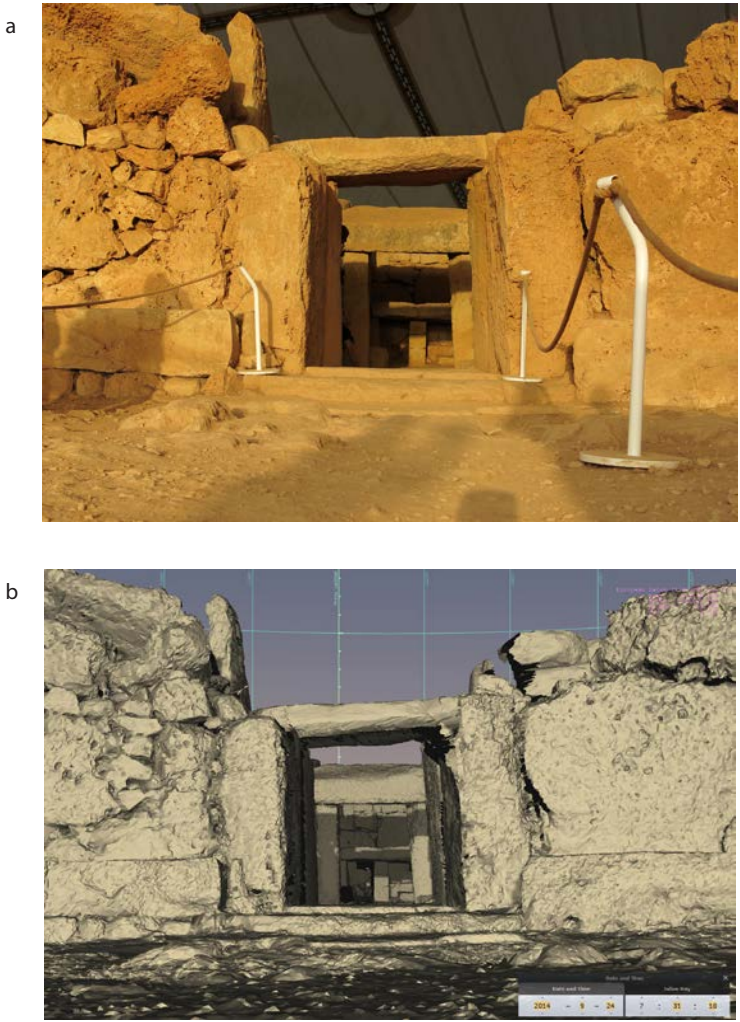

**FIGURE 4.** Sunrise in the southern temple of Mnajdra: (top) photograph taken on 24th September, 2014, by G. Zotti during the SEAC conference in Malta; (bottom) the same scene simulated at the same second of time in Stellarium 0.20.1. Compare the shadows cast along the stones into the inner sanctuary. Some differences stem from scanning errors (missing surfaces). Laser scan data courtesy of Heritage Malta.

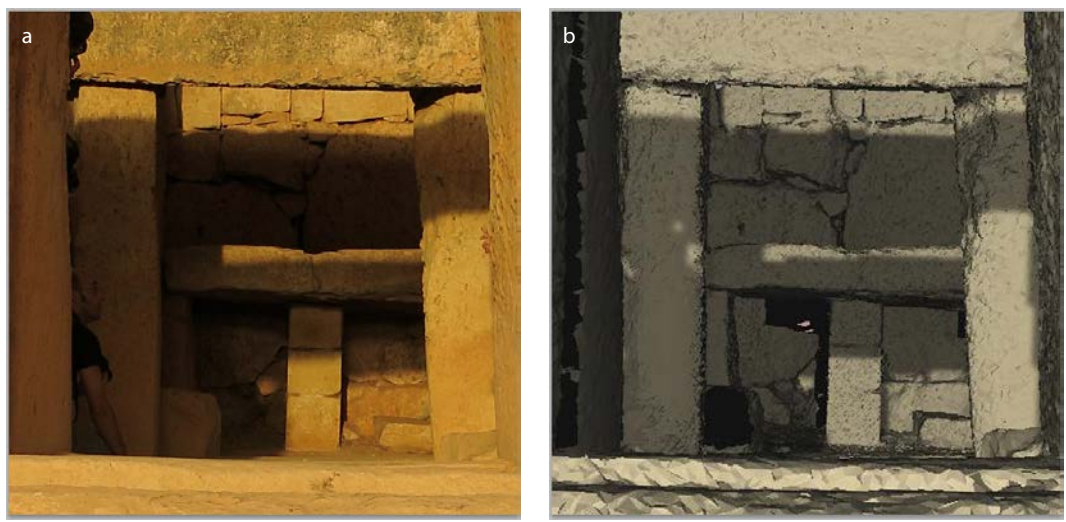

**FIGURE 5.** Enlarged central section of (a) the photograph and (b) the rendering from Figure 4. Some spots of light on the left vertical stone pillar are caused by data errors (holes) in the laser scan model. The placement of other shadows appears correct.

There are still noticeable differences in the scene appearance. However, most seem to relate to a slight lack of geometric resolution or missing geometry in the scan data, which leads to holes in the displayed walls which may then also cast wrong spots of light into shadowed areas. The natural Sun was dimmed by slight fog, so that the shadow contour of the ridge immediately at sunrise was not clearly discernible on the ground. Of course, the Sun's appearance as a disk means that the natural shadow does not have a sharp edge. In the simulation, we have tried several methods to soften or blur the hard shadow edges depending on distance to the geometry edge that casts the shadow, but some blocky appearance of the shadow edges can hardly be avoided for larger landscapes, due to technical restrictions of the shadow map texture size. Despite this, the light and shadow edges when the Sun finally cast discernible shadows into the temple are very closely simulated in the Scenery3D plugin (Figure 5). From this we gain confidence that the simulation should give similarly accurate results for light and shadow interaction for other dates, including Malta's temple era.

*Example 3: Chankillo*

The archaeological landscape of Chankillo, south of the Casma River near the coast of Peru, is a spectacular site dominated by two monumental structures (Ghezzi and Ruggles 2007; 2015). Observed from a particular corner in the remains of a building structure in the valley, 13 towers built along a ridge seem to mark the range of declinations which the rising Sun can reach during the year. To the west, a "fortified temple" overlooking the valley shows a very peculiar architecture of straight and circular walls. Site visits are costly, many viewpoints may have to be visited at the same time or over the course of the year to observe sunrises and sunsets, and the extended site, now mostly covered in sand, may even be too fragile to be explored by motorised vehicles. As such, the question was whether a 3D model could be used to explore the landscape in a walkthrough or low flyover mode to allow a simple illustration of the known interpretation and an investigation of further ideas around the archaeoastronomical concepts to be found in the landscape.

The immediate area around the 13 towers and the fortified temple has been recorded by aerial photography and LiDAR with a 1 m raster grid width, from which again a textured TIN was created in ArcGIS 10.4 and converted into the required georeferenced OBJ format. However, the mountains which form the eastern and northern horizon were not included. The SRTM 1" digital surface model (DSM) with its 1 arcsecond grid spacing (which is approximately equal to a 30 m grid spacing in the north–south direction; the actual east–west grid spacing depends on latitude) was used to compute a 3D mesh of terrain data where vertical accuracy and detail are not as important as in the near vicinity. However, high mountains may be visible to distances of almost 100 km and a fine-meshed TIN would easily become too large for the average PC. Given the need to include only higher parts of the mountains but not having details of valleys hidden by closer mountains which form the visible horizon, we created in the GIS an "area viewshed", an estimate of the terrain visible from any point in the central area of interest, and only then created the TIN from this visible part of the landscape. This reduced the triangle count significantly, but of course also limited the scope of the model, so that it could be viewed only from the valley area around the 13 towers, from the area of the fortified temple and not too high above ground. Even so, the highest-resolution model consisted of over 14 million triangles. The 2016 notebook PC can display this model including shadow simulation with over 10 frames per second (fps) on a wide 3440 × 1440 display in perspective mode. Using other projections, such as fisheye or cylindrical, again requires rendering the model on six sides of a foreground "cubemap" which is then re-projected into the scene, which in this case dropped the frame rate to below 3 fps while moving. This is too slow to be used comfortably interactively, but it can still be used for illustrations like those shown in Figures 6 and 7. With a setting of "lazy cubemap re-rendering" we can however look around with a higher frame rate from a fixed point, because the foreground needs only to be re-created when we are moving, or when the illumination changes. For even more detail, laser scans or image-based models of the temples and towers could still be added to the scenery, but this may require a more powerful graphics card with more memory than that in the laptop used here.

The very large extent of this model, however, shows a noteworthy limitation currently present in the landscape modelling's original intent (Zotti and Neubauer 2012), which is that it was intended to visualise a small piece of terrain with some archaeological or architectural artefact to be investigated in three dimensions, surrounded by a static representation (photographic, surveyed or artificial rendering) of the far-away mountains that does not perceivably shift when the observer moves around the site by a few metres. The central 3D scenery is currently modelled in Cartesian coordinates on a plane tangential to the Earth's surface and slightly rotated at loading time around the vertical axis to compensate for the local deviation of the survey grid axes (usually UTM) from the cardinal directions caused by meridian convergence (Zotti 2015). It therefore does not correct the scene for the Earth's curvature. In the case of this site, the farthest mountains seen from the "fortified temple" are somewhat short of 100 km to the north and 70 km to the east. We can estimate the error that results from planar modelling: for an observer at sea-level on the curved Earth, a mountain peak 1000 m high sinks below the mathematical horizon at a distance of about 113 km. In the planar model, this mountain in the same distance appears to be about 0.51°, or about a full solar diameter, high. In addition, the local azimuth of the horizontal axis in UTM coordinates deviates from the east–west direction depending on geographical latitude and distance from the UTM zone's central meridian. This means that azimuths of far mountain peaks seen in the planar model also deviate slightly.

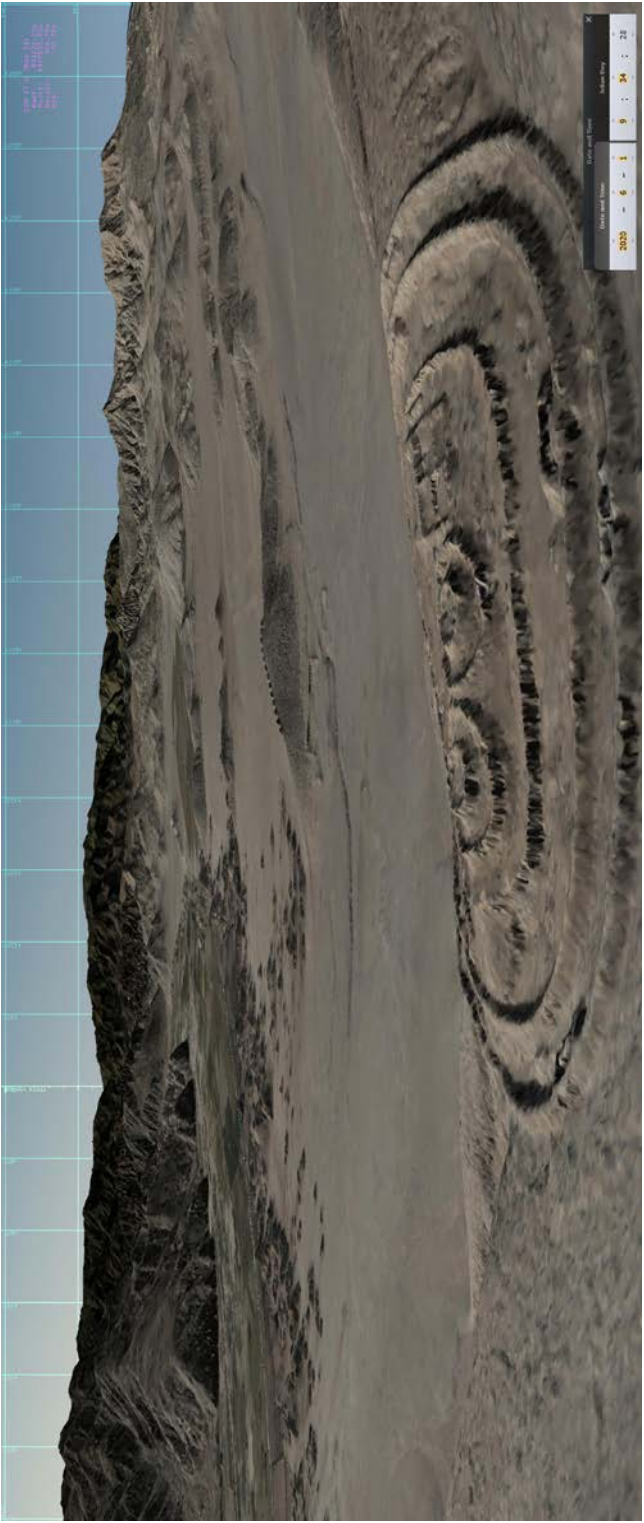

**FIGURE 6.** A simulated low-altitude aerial view overlooking the Chankillo 3D landscape model in Stellarium. The "fortified temple" is in the foreground, while the ridge with 13 towers and the "observation building" in front are visible in the valley below. The user can explore the whole archaeological area and inspect or discover new viewpoints of archaeoastronomical interest. The scene includes mountains almost 70 km to the east (LiDAR data courtesy Iván Ghezzi and Clive Ruggles).

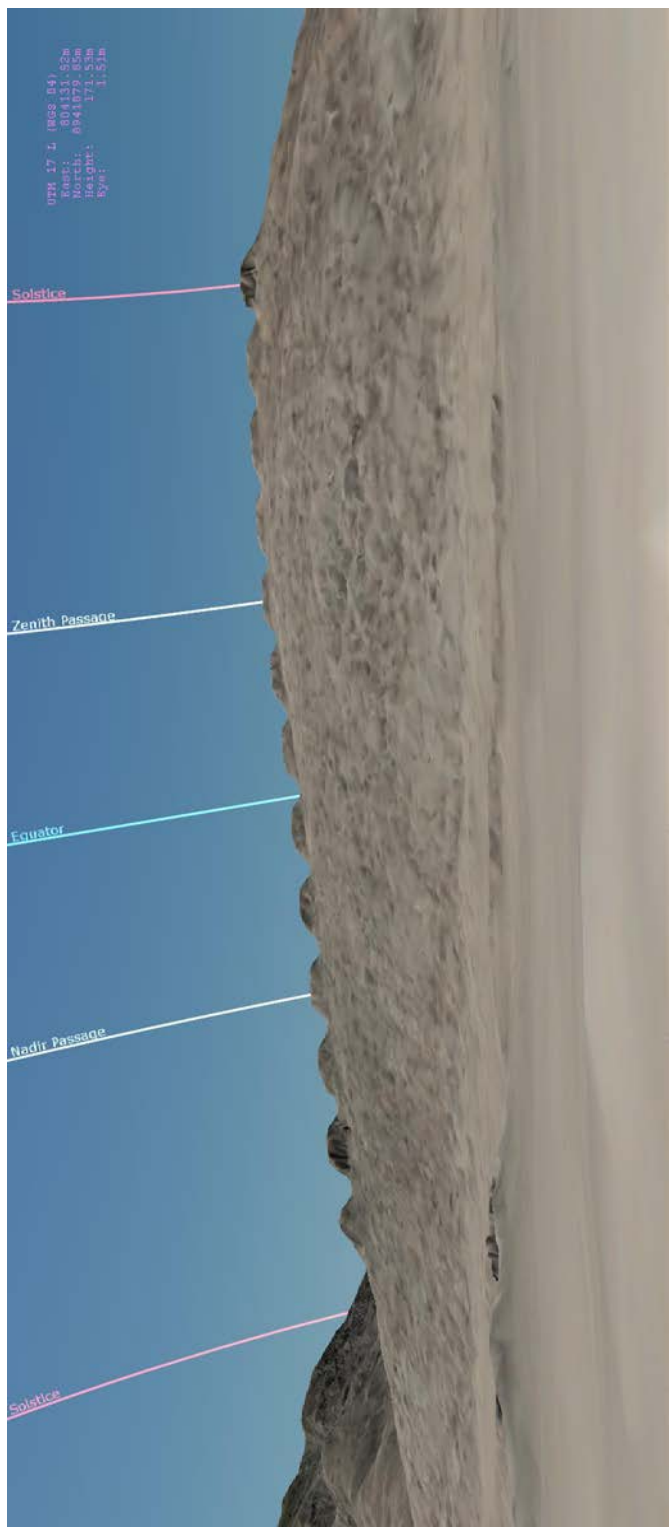

**FIGURE 7.** A Stellarium view from the "Western Observing Point" over the 13 towers in Chankillo visible in the LiDAR data, closely matching Figure 62.7a in Ghezzi and Ruggles (2015). The ArchaeoLines plugin is used to display diurnal tracks of the Sun at the solstices, celestial equator and zenith/nadir (antizenith) declinations in 300 BC. The leftmost tower has been smoothed in the LiDAR data and is somewhat higher in reality.

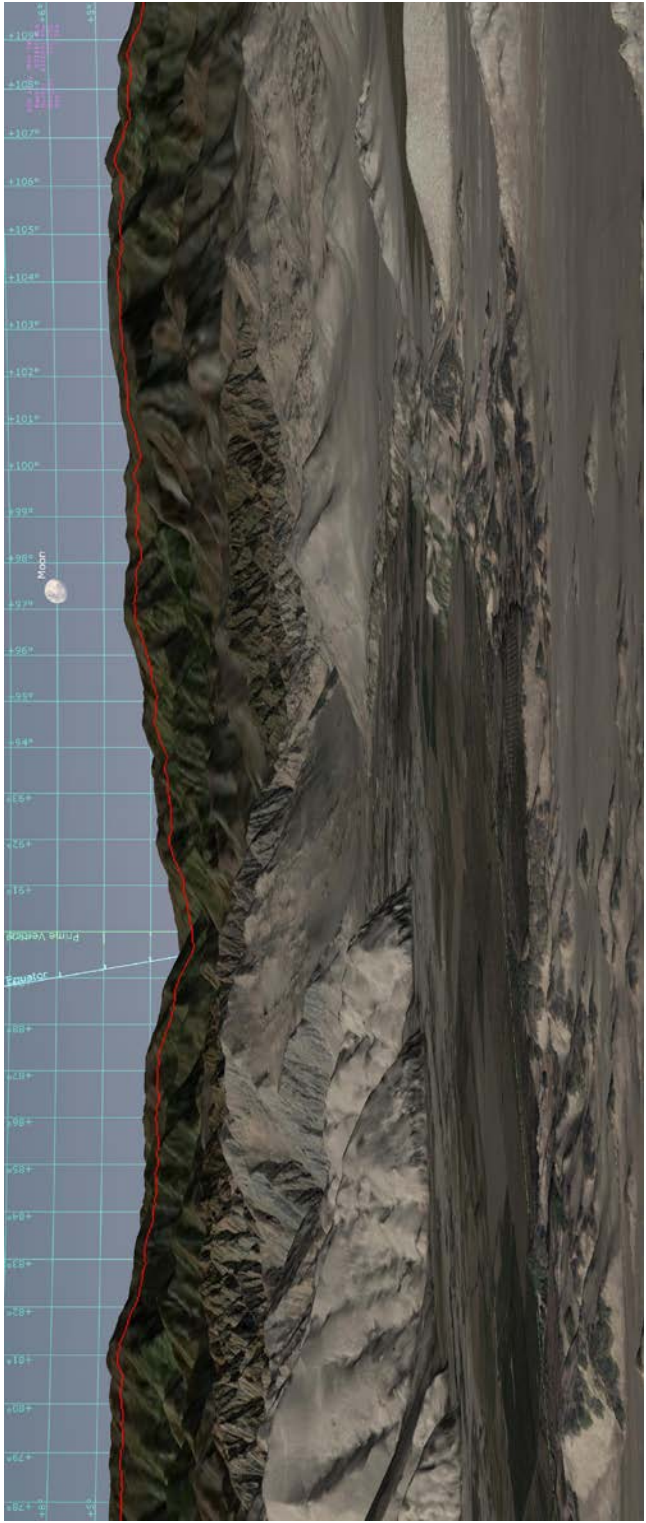

**FIGURE 8.** A close-up view towards the eastern horizon from the "Temple of the Pillars" in the Lidar-based 3D scenery of Chankillo. The red polygonal line was created from the same location with Horizon version 0.13c and is plotted over the model for clarity. The rising Moon and azimuthal grid (width 1°) are shown for comparison. This example shows an extreme case to illustrate the effect of not taking Earth's curvature into account. Distant mountains should preferably be included using a panorama that can be created with Horizon or actual survey data.

Another effect which counteracts the apparent depression of mountains is caused by the Earth's atmosphere, where terrestrial refraction can raise far-away mountains by several arcminutes, or more under exceptional atmospheric conditions. These effects are currently not modelled.

For comparison, the above-mentioned program Horizon (Smith 2020) was used to create a horizon panorama and polygon (from the same SRTM 1" data) as seen from the "Temple of the Pillars" (the right rectangular structure visible in Figure 6) in the fortified temple complex. The red horizon polygon created by the Horizon program is drawn after the 3D scenery to show where the "true" landscape horizon (actually, also only a digital approximation, but one that claims to correct for curvature!) should be visible when correcting for curvature (Figure 8). For distant mountains like in this example, the static panorama corrected for the Earth's curvature is clearly more accurate and this deviation should definitely not be neglected, but the impact on any proposition resulting from studying the virtual landscape in combination with celestial targets depends on the use case.

*Limitations*

The considerable size allowed for the OBJ 3D models in Stellarium invites the creation of large terrain models. As shown in the Chankillo example this approach has limitations caused by not modelling the curvature of Earth. Therefore, the recommended approach is to use a site of limited extent which is locally modelled in 3D and enclosed by a classical horizon panorama that should provide a representation of mountains far enough away so that their location does not shift perceivably when the observer moves inside the limited area of the site in question.

*Game-Like Interaction*

Another limitation of the 3D rendering module when compared to the typical user experience in today's first-person perspective 3D computer games is the static nature of the scene and lack of interaction possibilities with scene objects. The 3D plugin was developed strictly for the purpose of representing architectural and geological features, whereas in a computer game plants may move in the wind, animals or virtual characters controlled by algorithms or other simulation participants may move around and the user may interact with mobile scene parts. Such possibilities can be useful, for example to demonstrate the operation of historical astronomical observation instruments (Zotti and Mozaffari 2020). Such scenes call for the use of a game engine, a software toolkit which provides the building blocks required in modern computer games but usually has no sky model that fulfils even the most basic requirements for astronomical demonstrations. A simple sky simulation that includes a moving Sun and stellar sphere can be created with moderate effort (Zotti 2014). More recently, a combination of a simulation environment using the Unity game engine (Unity 2020) for the user interaction has been presented, along with Stellarium as a sky rendering engine which also provides the astronomically correct positions for shadow-casting light sources (Zotti 2020; Zotti *et al.* 2020). Depending on the final purpose of the game-like application, such a combination can reduce the high effort required for the development of an astronomically accurate sky simulator module for a game-like environment (Figure 9).

**FIGURE 9.** A virtual environment which allows the understanding and operation of historical astronomical observing instruments. The sky background and solar position for the Unity-based application are provided by Stellarium, controlled by its RemoteControl plugin.

**Stellarium for Sky Culture Research**

A very popular feature of Stellarium is the possibility of showing not only the common constellations officially approved by the International Astronomical Union (IAU – Delporte 1930), but also the constellations of a multitude of original "skycultures" of ethnological significance (more than 40 in version 0.20.2).

The Stellarium User Guide (Zotti and Wolf 2020b) describes the various ingredients for the current implementation of skycultures, which may consist of constellation names, constellation patterns ("stick figures"), artistic depictions of constellations, star names, planet names, names for deep-sky objects, constellation borders, "asterisms" (in the "Western" sense, those are non-official constellations, for example the Big Dipper as brightest part of Ursa Major), "ray helper lines" that aid orientation and a textual description. However, a few recent contributions and online discussions indicate that there is also some demand for the implementation of non-European skyculture concepts such as "dark constellations" (see, for example, Hamacher 2012; Fuller *et al.* 2014; Gullberg *et al.* 2020), single-star asterisms and the lunar stations observed in many Asian cultures.

The "Western" skyculture representing European scientific astronomy has a tradition of more than two millennia of star name development. The majority of names still comes from Ptolemy's star catalogue of the second century AD, in its translated Arabic and re-translated and often misspelled derivative forms. Only in recent years has the IAU, inheritor of the ancient Greek and later "Western" scientific tradition, started to assign authoritative names (IAU 2018). In some cases, new names were introduced from non-European cultures, so that some of them do not seem to fit the context of their mandatory IAU constellation. In other cases, traditional well-known (for example, Arabic) names have been simply assigned to other stars. For example, *Suhail*, in its solitary form, had been exclusively the traditional name of Canopus (α Car) in Arabic texts before it became the official name of λ Vel, probably due to an abbreviation from Allen's (1963 [1899], 74) identification as *al Suhail al Wazn*, one of three names mentioned by as-Sufi for three different stars without clear identification (Kunitzsch 1959, 209). Likewise, the official acknowledgement of *Gomeisa* for β CMi perpetuates Piazzi's misapplication (Kunitzsch 1959, 160) of the Arab name *Algomeisa*, which originally had been applied to Procyon (α CMi). Further cases, such as assigning the name *Okab* (eagle) to ζ Aql, replacing the complete traditional name of *Deneb el Okab* (tail of the eagle), cast doubts on this attempt to clarify names, because the "Star of the Eagle" had always been Altair (α Aql), the "Flying Eagle", in contrast to Vega (α Lyr), the "Swooping Eagle" (Kunitzsch and Smart 2006, 17, 43).

Stellarium predates these developments and includes star names from a multitude of sources, but to resolve further ambiguities, confusion or untraceable names, we started to add source information for the star names. The "Western" skyculture already includes this information (behind the screen), while most other skycultures are based on only one single source each. In upcoming versions we aim to let the user decide which sources for star names should be used by the program for each skyculture when more than one source is available. Outlines for further development have also been presented at SEAC 2019 (Zotti and Wolf, in press).

### Skycultures and Astrophysics

The analysis of historical astronomical records has become an important resource for research on long-term variability of stars (Fujiwara *et al.* 2004; Fujiwara and Yamaoka 2005; Hamacher and Frew 2010; Hamacher 2018), comet orbits (Neuhäuser *et al.* 2018), meteoritic events (Hamacher 2009), novae (Shara *et al.* 2007; Duerbeck 2008; Patterson *et al.* 2013; Miszalski *et al.* 2016; Shara *et al.* 2017), supernovae (Stephenson 1976; Clark and Stephenson 1977; Yau 1988; Stephenson and Green 2002; Green and Stephenson 2003; 2017; Zhou *et al.* 2018; Green 2019) and close interacting binary stars. Stellarium can be used to demonstrate transient phenomena like comets or "guest stars" recorded by historians and chroniclers (Zotti and Wolf 2020a; Hoffmann 2019b; Hoffmann *et al.* 2020; Hoffmann and Vogt 2020a).

Currently, Stellarium offers two possibilities to show "Bright Novae" and "Historical Supernovae" with two further plugins. The Bright Novae plugin works relatively simply: it displays novae that are not supernovae. Nowadays, the term "classical nova" is used only for eruptions in close binaries with amplitudes in brightness from 6 to 16 stellar magnitudes (mag). There are slow, moderately fast, fast and very fast novae, where the brightness declines by 2 mag over times ranging from 10 months to a few days. The close binaries (symbiotic stars or cataclysmic variables) that flare up as novae have been catalogued and their designation format follows the typical layout of variable star designations (i.e. two capital letters plus constellation name or "V" for "variable star" plus number plus constellation name). For example, the slowest nova eruption ever observed, which occurred in the cataclysmic binary V1500 Cyg, is named Nova Cygni 1975, whereas one of the most famous naked-eye novae, Nova Persei 1901, was a flare-up of the cataclysmic variable GK Per. The Stellarium user can find novae via a search tool by entering the designation of a nova event (e.g. Nova Persei 1901) or the common name of the binary system (e.g. GK Per).

The light curves of the declines of these recent novae in the past ~150 years are well known and Stellarium displays the nova event as a (bright) stellar object with a modelled rise and decline light curve derived from the observational data. The model is based on the time of decline by *n* magnitudes from the peak, where *n* is 2, 3, 6 and 9. In the case of very ancient historical novae, for example several Chinese guest star records from the past ~2500 years, no decline time is available. In these cases, the plugin uses typical values and a typical light curve (McLaughlin 1960, 585).

**FIGURE 10.** Supernova 1054 was reported in Chinese chronicles. Stellarium offers the appropriate context to show this event in combination with the Chinese constellations.

In contrast, the Historical Supernovae plugin has to cope with many uncertainties which should be treated more completely in upcoming versions: (1) in most of the ancient cases, the exact date of an appearance is not provided in historical sources and the duration and/or date of disappearance are completely missing; (2) the light curve cannot be reconstructed, because of a lack of historical observations, and it can be known only that something appeared (or disappeared); (3) it is not at all certain that we are dealing with a supernova rather than a nova or something else (Hoffmann and Vogt 2020b); and (4) typically, for ancient events, the position cannot be determined exactly because only the relevant constellation is preserved in the source.

A clear case is the supernova in 1054 AD, where the ancient reports all confirm a position within 2° of a well-known bright star. This way, the match of the historical event and the supernova remnant (Messier 1, the Crab Nebula) is certain (Mayall and Oort 1942) and we can map the historical transient object at its definite position. For a presentation in the cultural context of discovery and records in Chinese chronicles, the apparition can properly be presented in Stellarium with the constellations from the Chinese skyculture (Figure 10).

### *The Importance of Correct Skycultures*

If the skyculture in use for identifying likely candidates has been thoroughly prepared, it could also help directly in the identification of objects that have been observed and reported in the vicinity of named stars or parts of constellations of the respective skyculture. In this case, even astrophysicists without knowledge of the history of astronomical cultures could provide proper identifications of historical novae and supernovae. Therefore, future versions of Stellarium need to provide as many skycultures as possible with scrupulously investigated and proven references for the identification of stars, and with correctly defined validity in historical epoch and geographical region.

### *A Challenge for the History of Science*

A common problem of astrophysics is the short range of telescopic surveys, spanning roughly 200 years for systematic observations of sunspots and a few decades for certain types of stars such as optical transients. For studies of the long-term evolution of close variables (for example, cataclysmic and symbiotic binaries), supernova remnants, orbits of comets and even solar activities and space weather, it would be tremendously helpful to have access to data from the past two or three millennia of naked-eye observations. Medieval Christian and Islamic astronomers regarded the starry sky as immutable, and as such their work focused instead on tracking the positions of planets and there are only occasional records of bright transient phenomena (such as the supernova of AD 1006, which was noted in the annals of St Gallen monastery in Switzerland and some other European and Arab sources – Stephenson and Green 2002, 150, 167). In contrast, a large corpus of reported historical transients is preserved in Far Eastern chronicles (see e.g. Ho 1962; Xu *et al*. 2000; Pankenier *et al.* 2008). In Chinese astrological politics there was actually an emphasis on changes in the sky such as transient phenomena. Typically, the records are preserved only briefly in a schema like "date *x*, a new star appeared at asterism *a*" or "date *y*, a guest star was seen at asterism *b*; it lasted for *z* days". These transients could describe comets or meteors, but in cases without reported tail and movement, the nature of the appearance normally remains unclear.

In many cases, and as noted above in relation to supernovae, the positions of historical transients are described imprecisely. In some cases only the asterism is preserved, covering an area of several tens or hundreds of square degrees, while in others, the transient might be either a single star or a small pattern of two or three stars. Any historian who interprets the position (correctly), finds approximate coordinates in right ascension (RA) and declination (DEC) and finally publishes these in a table – with the best intentions, of course, along with text explaining that there are error ellipses of various size – will most likely be misunderstood by astrophysicists' quick reading. They will consider the table of point coordinates as the ultimate answer to the question of the position, and thus possibly derive wrong conclusions about which celestial object matches the position given in the historical record. Five papers were written to solve this problem, and one of us (SH) developed an unbiased method to compile a large corpus of historical records in order to make them accessible for modern research (Vogt *et al*. 2019; Hoffmann 2019b; Hoffmann *et al*. 2020; Hoffmann and Vogt 2020a, 2020b).

Stellarium, with its possibilities of visual representations – for example, highlighting areas in a map – could help as a tool here. Printed maps have only two spatial dimensions and thus either display a graphical representation or do not. This also applies for historical supernovae: in cases where there are alternative positions or identifications, a printed map cannot distinguish between these. Images, signs and diagrams can indicate two areas as both "this" AND "that" (e.g. a comet with long tail could be simultaneously in constellation areas *x* AND *y*) but they cannot clearly display a logical "OR" (*vel*, "inclusive or", for example overlapping candidate areas for target objects that are valid in both solutions) or even "XOR" (*aut*, "exclusive or", for example, candidate areas for the same text description which mutually exclude valid solutions). In cases of uncertainties concerning the identification of a historical record with one group of stars or another, the Stellarium map could easily be extended to display not only one star-like nova appearance and decline but additionally one or more area(s) in which alternative counterparts for the historically recorded apparition should be found; for examples, see maps in Hoffmann and Vogt (2020a, figs A5, A7, A9, A14 and A18). Mutually excluding alternatives, such as those in Hoffmann and Vogt (2020a, figs A8 and A12), could probably be displayed by blinking or in different colours. The difficulty in the interpretation of the simple table data of the twentieth century is impressively obvious in an article by Hoffmann (2019b, fig. 5) which shows the point coordinates of earlier scholars with attached error circles in comparison to the newly derived two alternative fields.

For example, "SN 386" is a transient which lasted for three months and the chronicle records only its appearance in the Chinese constellation of Nandou (the Southern Dipper), which covers half of the asterism in Sagittarius known popularly as the "Teapot". Thus, we know that there was a transient somewhere in an area of ~101 square degrees on the sphere. During the last few decades, three of the Sagittarius supernova remnants have been suggested as being the counterpart of this potential supernova (most recently, Zhou *et al.* 2018), and at least one further radio source appears possible. Furthermore, it is not at all certain that this transient really was a supernova: with a decline time of 2 or 3 mag within three months, it could also have been a classical nova. The very nature of the event is a hypothesis (Hoffmann and Vogt 2020c).

Thus, future improvements in Stellarium's supernovae plugin may aim at the possibilities of displaying areas in the sky where the transient event happened and adding information about suggested counterpart objects of any type. Displaying the area will also be tremendously helpful for scholars who want to use the historical data without being experts in interpreting it: it will thus allow the transfer of knowledge between academic disciplines.

**Discussion and Future Work**

This paper provides an overview of the current state of Stellarium (version 0.20.2) and highlights its features for cultural astronomy research. As with every tool, users should be aware of applicability limitations. Results derived with Stellarium are based on physically based models that have their own limits as described in the User Guide (Zotti and Wolf 2020b) or the respective sources.

The major improvements in astronomical accuracy have been listed above. As we are preparing version 0.21.0 of Stellarium (scheduled for March 2021) we are aware of minor open issues which keep us from declaring the software "finished", so the version number still starts with zero. (The number after that in the meantime corresponds to the year of release.) A user pointed us to a sign error in our implementation of nutation (based on a misprint in the original paper), bringing the times for the beginning of the seasons, which are defined by the Sun reaching geocentric ecliptical apparent longitudes of multiples of 90°, to within an encouraging few seconds of the reference solutions. The aberration of light however is still not yet implemented. This, or some other defect so far unidentified, causes errors of less than 1 minute for lunar occultation events. Another computational simplification from Stellarium's early days had caused the Moon to display part of its far side in prehistory. We were able finally to solve this, and other details of planet rotation, by implementing more accurate astronomical routines for version 0.21.0 (Urban and Seidelmann 2013).

A few common sources of error appear to result from misunderstandings in our common calendar. Historical chronology is used to start calendar epochs with a "Year One", usually referring to when a ruler came to the throne (the concept of zero cannot be expressed in Roman numerals). The Christian era is counted in years "Anno Domini"; the year before AD 1 (or 1 CE/Common Era, to use the religiously neutral alternative) is 1 BC (1 BCE. In contrast, astronomical counting of years includes a year zero, which is the same as the historical year 1 BC (Figure 11). If you want to visualise, for example, the sky in the year of Julius Caesar's assassination (44 BC), you must therefore switch Stellarium (and most other astronomy simulation programs) to the year −43.

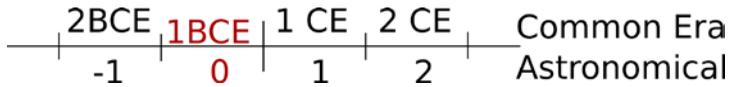

**FIGURE 11.** Counting the years at the beginning of the Common Era. Stellarium and most other astronomy simulations observe a year 0, which is named 1 BCE (or 1 BC) in historical chronology, and use negative year numbers for earlier dates. The numerical value of negative years is thus not the same as the same year given as BCE date.

Further, users should be aware that after Caesar's death his simple four-year rule was misapplied by the Romans and a leap day inserted every third year (Ginzel 1911, §189). It required another reform by the emperor Augustus, who had to omit three leap days before restarting the four-year sequence with the year AD 8 (8 CE, or year +8).

In the fourth century AD, the connection between the seasons and the date of Easter Sunday was codified in simple rules of calendar "computus". Over the centuries, however, it became evident that the Easter Moon and the seasons were coming adrift from fixed dates and there were several failed attempts to reform the calendar so that it would better synchronize with the seasons. Finally, in the sixteenth century Pope Gregory XIII decreed a reform that put the seasons back to the dates where they had been in the fourth century AD. Ten days were skipped in 1582 so that 4th October was followed by 15th October and the leap year rules were improved so that no leap year would be observed in years integrally divisible by 100, unless also divisible by 400. While Catholic countries soon adopted the reform and its rules which still hold today, Protestant countries ignored the papal decrees, which caused centuries of double-dating and confusion in several countries (Wilkins and Springett 1961, 414). Regardless of country and the correct date of adoption of the reformed calendar, most astronomy programs, and also Stellarium, conventionally use the Gregorian calendar for dates from 15th October, 1582, and the Julian calendar in the form finalised by Augustus for dates before that (Meeus 1998), especially also for dates of the late Roman Republic and early Roman Empire, so that displayed dates may be up to three days in error compared to dates possibly recorded by contemporary authors. When simulating or replaying historical observation records from Protestant countries, users may have to convert them to dates in the Gregorian calendar for correct simulation.

Given that the simple leap-year rule of Julius Caesar is applied for any date before 1583, months appear to shift out of seasons when turning back the time scale into centuries and millennia BC. For example, summer solstice in the astronomical year −2000 (=2001 BCE) was on a day labelled "10th July" in such a "proleptic" Julian calendar – far from "21st June", which most readers may be familiar with. If you for example want to set the day of solstice or equinox in the Neolithic or even earlier dates, it is recommended to use the displayed "ecliptical longitude of date" for the Sun, and keep in mind that the Sun is at 0° for spring equinox, 90° for summer solstice, 180° for autumn equinox, and 270° for winter solstice, because month names in the Julian calendar are for this and even earlier periods only misleading.

All these difficulties called for the addition of a new plugin to Stellarium v0.20.4 which directly supports various calendar systems.

The built-in VSOP87 ephemeris is recommended for use in the years −4000 to +8000 (4001 BC to AD 8000). As an analytic solution, it provides planetary positions also outside this range, where degraded accuracy must however be expected. This can be used to illustrate long-term effects like an approximation of the seasons in relation to the accurate orientation of the Earth's axis. It is even illustrative to follow the apparent motion of the Sun along the "ecliptic of date" to understand the implementation of the VSOP87 model. Going backwards from year −4000 to −4100 we can observe how the Sun is (artificially) deflected towards the ecliptic of astronomical epoch J2000, to remain there forever before −4100. On the other end of the recommended range, the same can be observed for years +8000 to +8100.

If planetary positions for a few millennia outside this date range are needed, there are instructions in the User Guide (Zotti and Wolf 2020b) which explain how to install and use the DE431 ephemeris file (Folkner *et al.* 2014), which provides data for the years −13,000 to +17,000, but without analytic extrapolation for years beyond this range (positions fall back to whatever VSOP87 provides). Extreme extrapolation will lead to a display of numerical nonsense, like the Moon running in a polar orbit around the year +78,000.

The long range of allowed years (−100,000 to +100,000) has been defined to study the principal effects of stellar proper motion. Stellar data currently include only the lateral motion components, and not radial velocity. Over millennia, stars can approach and recede, changing their apparent brightness in the process. Currently, Stellarium does not display effects such as these (De Lorenzis and Orofino 2018). Also, components of multiple stars do not show motion around a common centre of gravity, but fly apart at the rates controlled by the linear proper motion components recorded in the HIPPARCOS catalogue. In practice, however, only a handful of bright stars are noticeably affected. A better way to correct for stellar proper motions should therefore be worked out when the current star catalogues, based on HIPPARCOS (ESA 1997; Anderson and Francis 2012), Tycho 2 (Høg *et al.* 2000) and NOMAD (Zacharias *et al.* 2004) have undergone revision with data from the pending final version of the GAIA satellite observatory astrometry mission (Gaia Collaboration 2018; ESA 2020).

The simulation of minor bodies (asteroids, comets) depends on accurate orbital elements which have to be refreshed regularly for periodical objects. There should be one set of orbital elements for each object with an epoch (the time of exact validity for the osculating elements) near each perihelion.

Of course, Stellarium cannot fully replace archaeoastronomical fieldwork, just as it cannot replace the experience of standing under a natural cloudless night sky for children growing up in urban light-polluted environments. But it can augment the simulation and evaluation of past and present skies combined with archaeological data after site surveys, prospection, documentation, excavation and 3D mapping in ways that were unavailable just a few years ago.

Stellarium as a simulation engine offers a great wealth of applications for research in cultural astronomy, and it also has been used successfully in a large exhibition installa-tion (Zotti *et al.* 2017) and even for music performances (Fraietta 2019). Its open-source nature invites researchers – and in fact all users – to improve existing code and add new functionalities as they require, from more accurate atmospheric simulation to supporting the latest computerised telescopes, adding historical artwork from scanned copper print atlases or previously missing details to skyculture descriptions. Development is usually driven by the personal needs and ambitions of the voluntary developers and contributors. The possible extensions mentioned in this paper are not to be seen as announcements for the nearest future, but as directions of functionality in which computer programs dedicated to cultural astronomy research could further develop. Further suggestions for generally useful new features are usually welcome in the support forums, but not every suggestion can be implemented quickly by the small core team. Funded project collaboration, contributions in code or just pointers to better simulation models would be appreciated.

### Acknowledgements

We thank our anonymous reviewers who pointed out some issues for us to clarify.

Georg Zotti's work on Stellarium is in part supported by the Ludwig Boltzmann Institute for Archaeological Prospection and Virtual Archaeology (https://archpro.lbg.ac.at), which is based on an international cooperation of the Ludwig Boltzmann Gesellschaft (Austria), Amt der Niederösterreichischen Landesregierung (Austria), University of Vienna (Austria), TU Wien (Austria), ZAMG–Central Institute for Meteorology and Geodynamics (Austria), 7reasons (Austria), LWL–Federal state archaeology of Westphalia-Lippe (Germany), ArcTron 3D (Germany), NIKU–Norwegian Institute for Cultural Heritage (Norway) and Vestfold fylkeskommune–Kulturarv (Norway).